\documentstyle[floats,prl,aps]{revtex}
\begin{document}
\renewcommand{\thefootnote}{\fnsymbol{footnote}}
\draft
\title{\large\bf 
A new two-parameter integrable model of strongly correlated electrons
with quantum superalgebra symmetry} 

\author{Xiang-Yu Ge, Mark D. Gould, Yao-Zhong Zhang \footnote{
                                   E-mail: yzz@maths.uq.edu.au}
             and 
        Huan-Qiang Zhou \footnote {E-mail: hqz@maths.uq.edu.au}}

\address{      Department of Mathematics,University of Queensland,
		     Brisbane, Qld 4072, Australia}

\maketitle

\vspace{10pt}

\begin{abstract}
A new two-parameter integrable model with quantum superalgebra
$U_q[gl(3|1)]$ symmetry is proposed, which is an eight-state 
electron model with correlated single-particle and pair 
hoppings as well as uncorrelated triple-particle hopping.
The model is solved and the Bethe ansatz equations are obtained. 
\end{abstract}




\def\a{\alpha}
\def\b{\beta}
\def\d{\delta}
\def\e{\epsilon}
\def\g{\gamma}
\def\k{\kappa}
\def\l{\lambda}
\def\o{\omega}
\def\t{\theta}
\def\s{\sigma}
\def\D{\Delta}
\def\L{\Lambda}


\def\beq{\begin{equation}}
\def\eeq{\end{equation}}
\def\bea{\begin{eqnarray}}
\def\eea{\end{eqnarray}}
\def\ba{\begin{array}}
\def\ea{\end{array}}
\def\no{\nonumber}
\def\le{\langle}
\def\re{\rangle}
\def\lt{\left}
\def\rt{\right}

\newcommand{\reff}[1]{eq.~(\ref{#1})}

\vskip.3in
Integrable strongly correlated electron systems (see, e.g. \cite{Ess94}) 
have been an important subject of research activity (see e.g.
\cite{Kaw90,Bar95}).
In a recent paper \cite{Gou97}, we proposed an integrable correlated
electron model which is an eight-state version of the supersymmetric
$U$ model \cite{Bra95}. The latter has been extensively investigated by
many authors \cite{Bed95}. By eight-state we mean that at
a given lattice site $j$ of the length $L$
there are eight possible electronic states:
\bea
&&|0\re\,,~~~c_{j,1}^\dagger|0\re\,,~~~
  c_{j,2}^\dagger|0\re\,,~~~ c_{j,3}^\dagger|0\re,\no\\
&&c_{j,1}^\dagger c_{j,2}^\dagger|0\re\,,~~~ 
  c_{j,1}^\dagger c_{j,3}^\dagger|0\re\,,~~~ 
  c_{j,2}^\dagger c_{j,3}^\dagger|0\re\,,~~~ 
  c_{j,1}^\dagger c_{j,2}^\dagger c_{j,3}^\dagger|0\re\,,\label{states}
\eea
where $c_{j,\a}^\dagger$ ($c_{j,\a}$) denotes a fermionic creation
(annihilation) operator which creates (annihilates) an electron
of species $\a=1,\;2,\;3$ at
site $j$; these operators satisfy the anti-commutation relations given by
$\{c_{i,\a}^\dagger, c_{j,\b}\}=\d_{ij}\d_{\a\b}$.
The eight-state supersymmetric $U$ model contain one free parameter and
has Lie superalgebra $gl(3|1)$ as its symmetry.

In this letter, we promote the Lie superalgebra $gl(3|1)$ symmetry of
the model to the quantum superalgebra $U_q[gl(3|1)]$ symmetry, thus
giving a new integrable model of strongly correlated electrons with
two free parameters. We then solve the two-parameter model by the
coordinate space Bethe ansatz method and derive the Bethe ansatz
equations. 

The Hamiltonian for our new model on a periodic lattice reads
\bea
H(g,\k)&=&\sum _{j=1}^L H_{j,j+1}(g,\k),\no\\
H_{j,j+1}(g,\k)&=&-\sum_\a(c_{j,\a}^\dagger c_{j+1,\a}+{\rm h.c.})
  \exp\lt\{-\frac{1}{2}(\eta +\kappa) \sum_{\b(\neq\a)}n_{j+\theta(\b-\a),\b}
  -\frac {1}{2}(\eta-\kappa) \sum _{\b(\neq \a)}
  n_{j+1-\theta (\b-\a),\b}\rt.\no\\
   & &\lt.+\frac{\zeta}{2}
  \sum_{\b\neq\g(\neq\a)}(n_{j,\b}n_{j,\g}
  +n_{j+1,\b} n_{j+1,\g})\rt\}
-\frac{\sinh \kappa}{2\sinh \kappa(g+1)}
\sum_{\a\neq\b\neq\g}(c_{j,\a}^\dagger c_{j,\b}^\dagger 
  c_{j+1,\b}c_{j+1,\a}+{\rm h.c.})\no\\
 & &\exp\lt\{-(\frac{\xi}{2}-{\rm sign}(\g-2)\kappa)n_{j,\g}-(\frac {\xi}{2}+
  {\rm sign}(\g-2)\kappa)n_{j+1,\g}\rt\}\no\\
& &-\frac{2\cosh \kappa \sinh^2 \kappa}{\sinh \kappa(g+1)\sinh \kappa
(g+2)}\lt(c_{j,1}^\dagger c_{j,2}^\dagger 
  c_{j,3}^\dagger c_{j+1,3} c_{j+1,2} c_{j+1,1}+{\rm h.c.}\rt)\no\\
& & +e^{\kappa g}\; n_{j}+e^{-\kappa g}\;
  n_{j+1}-\frac{\sinh \kappa}{2\sinh \kappa (g+1)}\sum_{\a\neq\b}
  (n_{j,\a}n_{j,\b}+n_{j+1,\a}n_{j+1,\b})\no\\
& &+\frac{2\cosh \kappa (g+1)\sinh^2 \kappa}{\sinh \kappa (g+1)\sinh
  \kappa (g+2)}(n_{j,1}n_{j,2}n_{j,3}+n_{j+1,1}n_{j+1,2}
  n_{j+1,3}),\label{hamiltonian}
\eea
where $g,~\k$ are two free parameters, $n_j=n_{j,1}+n_{j,2}+n_{j,3}$ with
$n_{j,\a}=c_{i,\a}^\dagger c_{j,\a}$ being the number operator
for the electron of species $\a$ at site $j$,
$\t(\b-\a)$ is a step function of $(\b-\a)$ and
\beq
\eta=-\ln\frac{\sinh \k g}{\sinh\k(g+1)},~~~~\zeta=\frac{1}{2}
  \ln\frac{\sinh^2\k(g+1)}{\sinh\k g\,\sinh\k(g+2)},~~~~
  \xi=-\ln\frac{\sinh\k g}{\sinh\k(g+2)}.
\eeq

The model contains correlated single-particle and pair hoppings,
uncorrelated triple-particle hopping, and two- and three-particle
on site interactions. As is seen below, the chemical potential terms
are essential for the model to have the quantum superalgebra symmetry.
It should be noted that for the periodic lattice chemical potentials
can be dropped from the Hamiltonian without affecting the integrability.
However, they play an essential role for an open lattice.
As we have demonstrated in \cite{Zha97}, without these terms, the
boundary system would not be solvable by the coordinate Bethe ansatz 
method for a large class of integrable boundary conditions 
\cite{Zha97,Arn97}.

Some remarks are in order. For the choice of
$\k=0$, the above model becomes the eight-state supersymmetric $U$ model
proposed in \cite{Gou97}, whereas in the limit of $\eta=\k$ (first
discarding the chemical potential terms in the Hamiltonian
and then taking the limit), it
reduces to the electron model introduced in \cite{Bar91}:
\beq
H(g)=-\sum _{j=1}^L
  \sum_\a(c_{j,\a}^\dagger c_{j+1,\a}+{\rm h.c.})
  \;exp\lt\{-\eta \sum_{\b(\neq\a)}n_{j+\theta(\b-\a),\b}\rt\} \label
  {bariev}
\eeq
whose integrability is established in \cite{Zho97}.

The model defined by (\ref{hamiltonian}) is $U_q[gl(3|1)]$ supersymmetric
and is exactly solvable on the one-dimensional periodic lattice.
This is because the local Hamiltonian $H_{j,j+1}(g,\k)$
is actually derived through the quantum inverse
scattering method  using a $U_q[gl(3|1)]$ invariant R-matrix. 
To show this, we denote the generators of $U_q[gl(3|1)]$ by $E^\mu_\nu,~~
\mu,\nu=1,2,3,4$ with grading $[1]=[2]=[3]=0,~[4]=1$. In a typical 
8-dimensional representation $V(\L)$ of $U_q[gl(3|1)]$, 
the highest weight $\L=(0,0,0|g)$ itself of the
representation depends on the free parameter $g$, thus giving rise to
a one-parameter family of inequivalent irreps. Let $\{|x\re\}_{x=1}^8$
denote an orthonormal basis with $|1\re, |5\re, |6\re, |7\re$ even
(bosonic) and $|2\re, |3\re, |4\re, |8\re$ odd (fermionic). Then
the simple generators $\{E^\mu_\mu\}_{\mu=1}^4$ and $\{E^\mu_{\mu+1},\;
E^{\mu+1}_\mu\}_{\mu=1}^3$ are $8\times 8$ supermatrices of the form
\bea
&&E^1_2=|3\re \le 4|+|5\re\le 6|,~~~E^2_1=|4\re\le 3|+|6\re\le 5|,~~~
  E^1_1=-|4\re\le 4|-|6\re\le 6|-|7\re\le 7|-|8\re\le 8|,\no\\
&&E^2_3=|2\re\le 3|+|6\re\le 7|,~~~
  E^3_2=|3\re\le 2|+|7\re\le 6|,~~~
  E^2_2=-|3\re\le 3|-|5\re\le 5|-|7\re\le 7|-|8\re\le 8|,\no\\
&&E^3_4=\sqrt{[g]_q}\,|1\re\le 2|+\sqrt{[g+1]_q}\,(|3\re\le 5|+|4\re\le 6|)
   +\sqrt{[g+2]_q}\,|7\re\le 8|,\no\\
&&E^4_3=\sqrt{[g]_q}\,|2\re\le 1|+\sqrt{[g+1]_q}\,(|5\re\le 3|+|6\re\le 4|)
   +\sqrt{[g+2]_q}\,|8\re\le 7|,\no\\
&&E^3_3=-|2\re\le 2|-|5\re\le 5|-|6\re\le 6|-|8\re\le 8|,\no\\
&&E^4_4=g\,|1\re\le 1|+(g+1)\,\left (|2\re\le 2|+|3\re\le 3|
   |4\re\le 4|\right )
  +(g+2)\,(|5\re\le 5|+|6\re\le 6|+|7\re\le 7|)
  +(g+3)\,|8\re\le 8|,\label{matrix}
\eea
where $[x]_q=(q^x-q^{-x})/(q-q^{-1})$.

$U_q(gl(3|1))$ is a graded Hopf algebra with coproduct given by
\bea
\D(E^\mu_\mu)&=&I\otimes E^\mu_\mu+E^\mu_\mu \otimes I,~~~~ \mu=1,2,3,4,\no\\
\D(E^\mu_{\mu+1})&=&E^\mu_{\mu+1} \otimes q^{\frac{1}{2}(E^\mu_\mu-
   (-1)^{[\mu+1]}E^{\mu+1}_{\mu+1})}+
   q^{-\frac{1}{2}(E^\mu_\mu-(-1)^{[\mu+1]}E^{\mu+1}_{\mu+1})}
   \otimes E^\mu_{\mu+1},\no\\
\D(E^{\mu+1}_{\mu})&=&E^{\mu+1}_{\mu} \otimes q^{\frac{1}{2}(E^\mu_\mu-
   (-1)^{[\mu+1]}E^{\mu+1}_{\mu+1})}+
   q^{-\frac{1}{2}(E^\mu_\mu-(-1)^{[\mu+1]}E^{\mu+1}_{\mu+1})}
   \otimes E^{\mu+1}_{\mu},~~~~\mu=1,2,3.
\eea

Associated with the 8-dimensional representation, there is 
a $U_q[gl(3|1)]$-invariant R-matrix
which satisfies the graded Yang-Baxter equation. The R-matrix
is given by
\beq
\check{R}(u)=\check{P}_1+\le 2g \re\check{P}_2
  +\le 2g \re \le 2g+2 \re\check{P}_3
  +\le 2g \re \le 2g+2 \re \le 2g+4 \re\check{P}_4,
 \label{rational-R}
\eeq
where $\le a\re=(1-q^{u+a})/(q^u-q^a)$ and
$\check{P}_k:~ \check{P}_k[V(\L)\otimes V(\L)]=V(\L_k),~k=1,2,3,4$,
are four projection operators which now going to
construct. $\check{P}_1$ and $\check{P}_4$ are
\beq
\check{P}_1=\sum_{k=1}^8 |\Psi_k^1\re\le\Psi_k^1|,~~~~~~
  \check{P}_4=\sum_{k=1}^8|\Psi_k^4\re\le\Psi_k^4|.\label{projector1}
\eeq
Throughout this paper,
\bea
&&\le\Psi^a_k|=\left (|\Psi^a_k\re\right )^\dagger,~~~~
  \left (|x\re\otimes |y\re\right )^\dagger=(-1)^{[|x\re][|y\re] }
  \le y|\otimes \le x|\label{dual1}
\eea
with $[|x\re]=0$ for even (bosonic) $|x\re$ and $[|x\re]=1$ for odd
(fermionic) $|x\re$. $|\Psi^1_k\re,~ 
|\Psi_k^4\re,~k=1,2,\cdots,8$ are given by
\bea
|\Psi^1_1\re&=&|1\re\otimes |1\re,\no\\
|\Psi^1_i\re&=&\frac{1}{\sqrt{q^{g}+q^{-g}}}(q^{\frac{g}{2}}
  |i\re\otimes |1\re+q^{-\frac{g}{2}}|1\re\otimes |i\re),~~~i=2,3,4,\no\\
|\Psi^1_i\re&=&\frac{1}{\sqrt{(q^{g}+q^{-g})[2g+1]_q}}[\sqrt{[g+1]_q}
  (q^{g}|i\re\otimes |1\re +q^{-g}|1\re\otimes |i\re)\no\\
& &  +\sqrt{[g]_q}
  (q^{\frac{1}{2}}|2\re\otimes |i-2\re-
  q^{-\frac{1}{2}}|i-2\re\otimes |2\re)],~~~i=5,6,\no\\
|\Psi^1_7\re&=&\frac{1}{\sqrt{(q^{g}+q^{-g})[2g+1]_q}}[\sqrt{[g+1]_q}
  (q^{g}|7\re\otimes |1\re +q^{-g}|1\re\otimes |7\re)+\sqrt{[g]_q}
  (q^{\frac{1}{2}}|3\re\otimes |4\re-
  q^{-\frac{1}{2}}|4\re\otimes |3\re)],\no\\
|\Psi^1_8\re&=&\frac{1}{\sqrt{(q^{g}+q^{-g})(q^{g+1}+q^{-g-1})[2g+1]_q}}
   [\sqrt{[g]_q}(q^{-\frac{g}{2}+1}|2\re\otimes |7\re 
  +q^{\frac{g}{2}-1}|7\re\otimes |2\re
  +q^{\frac{g}{2}+1}|5\re\otimes 4\re\no\\
& &+q^{-\frac{g}{2}-1}|4\re\otimes |5\re
 -q^{-\frac{g}{2}}|3\re\otimes |6\re-q^{\frac{g}{2}}|6\re\otimes |3\re)
  +\sqrt{[g+2]_q}(q^{\frac{3g}{2}}|8\re\otimes 
  |1\re+q^{-\frac{3g}{2}}|1\re\otimes |8\re)],\no\\
|\Psi^4_1\re&=&\frac{1}{\sqrt{(q^{g+1}+q^{-g-1})(q^{g+2}+
  q^{-g-2})[2g+3]_q}}[\sqrt{[g+2]_q}(q^{-\frac{g}{2}-2}
  |7\re\otimes |2\re
  -q^{\frac{g}{2}+2}|2\re\otimes |7\re\no\\
& &+q^{-\frac{g}{2}}|5\re\otimes |4\re-
  q^{\frac{g}{2}}|4\re\otimes |5\re
  +q^{\frac{g}{2}+1}|3\re\otimes |6\re-
  q^{-\frac{g}{2}-1}|6\re\otimes |3\re)\no\\
& &  +\sqrt{[g]_q}(q^{\frac{3g}{2}+3}|1\re\otimes |8\re
  -q^{-\frac{3g}{2}-3}|8\re\otimes |1\re)],\no\\
|\Psi^4_2\re&=&\frac{1}{\sqrt{(q^{g+2}+q^{-g-2})[2g+3]_q}}
  [\sqrt{[g+1]_q}(q^{-g-2}|8\re\otimes |2\re +q^{g+2}|2\re\otimes 
  |8\re)\no\\
& &+\sqrt{[g+2]_q}(q^{\frac{1}{2}}|5\re\otimes |6\re-
  q^{-\frac{1}{2}}|6\re\otimes |5\re)],\no\\
|\Psi^4_i\re&=&\frac{1}{\sqrt{(q^{g+2}+q^{-g-2})[2g+3]_q}}
  [\sqrt{[g+1]_q}(q^{-g-2}|8\re\otimes |i\re +q^{g+2}|i\re\otimes 
  |8\re)\no\\
& & +\sqrt{[g+2]_q}(q^{\frac{1}{2}}|i+2\re\otimes |7\re-
  q^{-\frac{1}{2}}|7\re\otimes |i+2\re)],~~~i=3,4,\no\\
|\Psi^4_i\re&=&\frac{1}{\sqrt{q^{g+2}+q^{-g-2}}}(-q^{-\frac{g}{2}-1}
  |8\re\otimes |i\re+q^{\frac{g}{2}+1}|i\re\otimes |8\re),~~~i=5,6,7,\no\\
|\Psi^4_8\re&=&|8\re\otimes |8\re.\label{basis1}
\eea

The construction of the projector $\check{P}_2$ is a bit involved. 
We first construct the
basis vectors for $V(\L_2)$. With the help of the coproduct formulae
and after some algebraic calculations, we get
\bea 
|\Psi^2_1\re&=&\frac{1}{\sqrt{q^{g}+q^{-g}}}(-q^{-\frac{g}{2}}
  |2\re\otimes |1\re+q^{\frac{g}{2}}|1\re\otimes |2\re),\no\\
|\Psi^2_2\re&=&\frac{1}{\sqrt{q^{g}+q^{-g}}}(-q^{-\frac{g}{2}}
  |3\re\otimes |1\re+q^{\frac{g}{2}}|1\re\otimes |3\re),\no\\
|\Psi^2_3\re&=&|2\re\otimes |2\re,\no\\
|\Psi^2_4\re&=&\frac{1}{\sqrt{q^{g}+q^{-g}}}(-q^{-\frac{g}{2}}
  |4\re\otimes |1\re+q^{\frac{g}{2}}|1\re\otimes |4\re),\no\\
|\Psi^2_5\re&=&\frac{1}{\sqrt{q+q^{-1}}}(q^{-\frac{1}{2}}|2
  \re\otimes |3\re+q^{\frac{1}{2}}|3\re\otimes |2\re),\no\\
|\Psi^2_6\re&=&|3\re\otimes |3\re,\no\\
|\Psi^2_7\re&=&\frac{1}{\sqrt{q+q^{-1}}}(q^{-\frac{1}{2}}|2\re\otimes 
  |4\re+q^{\frac{1}{2}}|4\re\otimes |2\re),\no\\
|\Psi^2_8\re&=&\frac{1}{\sqrt{q+q^{-1}}}(q^{-\frac{1}{2}}|3\re\otimes 
  |4\re+q^{\frac{1}{2}}|4\re\otimes |3\re),\no\\
|\Psi^2_9\re&=&|4\re\otimes |4\re,\no\\
|\Psi^2_{10}\re&=&\frac{1}{\sqrt{q^{g+1}+q^{-g-1}}}(-q^{-\frac{g}{2}-1}
  |2\re\otimes |5\re+q^{\frac{g}{2}+1}|5\re\otimes |2\re),\no\\
|\Psi^2_{11}\re&=&\frac{1}{\sqrt{q^{g+1}+q^{-g-1}}}(-q^{-\frac{g}{2}-1}
  |3\re\otimes |5\re+q^{\frac{g}{2}+1}|5\re\otimes |3\re),\no\\
|\Psi^2_{12}\re&=&\frac{1}{\sqrt{q^{g+1}+q^{-g-1}}}(-q^{-\frac{g}{2}-1}
  |2\re\otimes |6\re+q^{\frac{g}{2}+1}|6\re\otimes |2\re),\no\\
|\Psi^2_{13}\re&=&\frac{1}{\sqrt{q^{g+1}+q^{-g-1}}}(-q^{-\frac{g}{2}-1}
  |4\re\otimes |6\re+q^{\frac{g}{2}+1}|6\re\otimes |4\re),\no\\
|\Psi^2_{14}\re&=&\frac{1}{\sqrt{q^{g+1}+q^{-g-1}}}(-q^{-\frac{g}{2}-1}
  |3\re\otimes |7\re+q^{\frac{g}{2}+1}|7\re\otimes |3\re),\no\\
|\Psi^2_{15}\re&=&\frac{1}{\sqrt{q^{g+1}+q^{-g-1}}}(-q^{-\frac{g}{2}-1}
  |4\re\otimes |7\re+q^{\frac{g}{2}+1}|7\re\otimes |4\re),\no\\
|\Psi^2_{16}\re&=&\frac{1}{\sqrt{(q^{g}+q^{-g})[2g+1]_q}}
  [\sqrt{[g]_q}(q^{g+\frac{1}{2}}|2\re\otimes |3\re 
  +q^{-g-\frac{1}{2}}
  |3\re\otimes |2\re)+\sqrt{[g+1]_q}(|1\re\otimes |5\re-
  |5\re\otimes |1\re)],\no\\
|\Psi^2_{17}\re&=&\frac{1}{\sqrt{(q^{g}+q^{-g})[2g+1]_q}}[\sqrt{[g]_q}
  (q^{g+\frac{1}{2}}|2\re\otimes |4\re 
  +q^{-g-\frac{1}{2}}|4\re\otimes 
  |2\re)+\sqrt{[g+1]_q}(|1\re\otimes |6\re-|6\re\otimes |1\re)],\no\\
|\Psi^2_{18}\re&=&\frac{1}{\sqrt{(q^{g}+q^{-g})[2g+1]_q}}[\sqrt{[g]_q}
  (q^{g+\frac{1}{2}}|3\re\otimes |4\re +q^{-g-\frac{1}{2}}|4\re\otimes 
  |3\re)
  +\sqrt{[g+1]_q}(|1\re\otimes |7\re-|7\re\otimes |1\re)],\no\\
|\Psi^2_{19}\re&=&\frac{1}{\sqrt{(q+q^{-1})(q^{g+1}+q^{-g-1})}}
  (q^{\frac{g}{2}+1}
  |6\re\otimes |3\re -q^{-\frac{g}{2}-1}|3\re\otimes |6\re
  +q^{\frac{g}{2}}
  |5\re\otimes |4\re-q^{-\frac{g}{2}}|4\re\otimes |5\re)],\no\\
|\Psi^2_{20}\re&=&\frac{1}{\sqrt{(q+q^{-1})(q^{g+1}+q^{-g-1})}}
  (q^{\frac{g}{2}+1}|7\re\otimes |2\re -q^{-\frac{g}{2}-1}|2\re\otimes 
  |7\re
  +q^{\frac{g}{2}}|6\re\otimes |3\re-
  q^{-\frac{g}{2}}|3\re\otimes |6\re)],\no\\
|\Psi^2_{21}\re&=&\frac{1}{\sqrt{(q^{g+1}+q^{-g-1})[2g+3]_q}}
  [\sqrt{[g+1]_q}(q^{-\frac{1}{2}}|6\re\otimes |5\re- q^{\frac{1}{2}}
  |5\re\otimes |6\re)\no\\
& &  +\sqrt{[g+2]_q}(q^{g+1}|8\re\otimes |2\re+q^{-g-1}
  |2\re\otimes |8\re)],\no\\
|\Psi^2_{22}\re&=&\frac{1}{\sqrt{(q^{g+1}+q^{-g-1})[2g+3]_q}}
  [\sqrt{[g+1]_q}(q^{-\frac{1}{2}}|7\re\otimes |5\re- q^{\frac{1}{2}}
  |5\re\otimes |7\re)\no\\
& &  +\sqrt{[g+2]_q}(q^{g+1}|8\re\otimes 
  |3\re+q^{-g-1}|3\re\otimes |8\re)],\no\\
|\Psi^2_{23}\re&=&\frac{1}{\sqrt{(q^{g+1}+q^{-g-1})[2g+3]_q}}
  [\sqrt{[g+1]_q}(q^{-\frac{1}{2}}|7\re\otimes |6\re- q^{\frac{1}{2}}
  |6\re\otimes |7\re)\no\\
& &  +\sqrt{[g+2]_q}(q^{g+1}|8\re\otimes |4\re+q^{-g-1}
  |4\re\otimes |8\re)],\no\\
|\Psi^2_{24}\re&=&\frac{1}{\sqrt{(q^{g}+q^{-g})(q^{g+1}
  +q^{-g-1})[2g+1]_q}}
  [\sqrt{[g]_q}(q^{\frac{g}{2}+1}|2\re\otimes |7\re 
  -q^{-\frac{g}{2}-1}|7\re\otimes |2\re\no\\
& &+q^{\frac{3g}{2}+1}|5\re\otimes |4\re-
  q^{-\frac{3g}{2}-1}|4\re\otimes |5\re
  -q^{\frac{g}{2}}|3\re\otimes |6\re-
  q^{-\frac{g}{2}}|6\re\otimes |3\re)\no\\
& &  +\sqrt{[g+2]_q}(-q^{\frac{g}{2}}|8\re\otimes |1\re+
  q^{-\frac{g}{2}}|1\re\otimes |8\re)].\label{basis2}
\eea
However, this basis for $V(\L_2)$ is not orthogonal, that is
$\le\Psi^2_l|\Psi^2_m\re\neq\d_{lm}$ for some $l,\;m$. To make it
orthogonal, we define a metric matrix $g_{lm}$ by
\beq
g_{lm}=\le \Psi^2_l|\Psi^2_m\re,~~~~~~l,m=1,2,\cdots,24,\label{metric}
\eeq
where $\le\Psi^2_k|$ are defined by (\ref{dual1}). We then form a
dual basis by means of the metric matrix, with basis vectors given by
\beq
\le\Psi^{2,l}|=\sum_{m=1}^{24}g^{lm}\le\Psi^2_m|,\label{dual2}
\eeq
where $g^{lm}\equiv (g^{-1})_{lm}$ is the inverse of the metric matrix
$g_{lm}$.
Then by construction, $\le\Psi^{2,l}|\Psi^2_m\re=\d_{lm}$ for all
$l,\;m$, which implies that $\le\Psi^{2,l}|$ are orthogonal to
$|\Psi^2_m\re$ for all $l,\;m$. Thus the projector $\check{P}_2$ is given by
\beq
\check{P}_2=\sum_{k=1}^{24}\;|\Psi^2_k\re\le\Psi^{2,k}|.\label{projector2} 
\eeq
Finally the projector $\check{P}_3$ is obtained through the relation
$\check{P}_1+\check{P}_2+\check{P}_3+\check{P}_4=1$.

On the $L$-fold tensor product space $V\otimes V\otimes \cdots\otimes V$
we denote $\check{R}(u)_{j,j+1}=
I^{\otimes(j-1}\otimes\check{R}(u)\otimes I^{\otimes(L-j-1)}$, and
define the local Hamiltonian by
\beq
H^{\rm R}_{j,j+1}(g,q)=\left .\frac{d}{du}\check{R}_{j,j+1}
   (u)\right |_{u=0}
\eeq
We make the identifications:
\bea
&&|1\re=|0\re\,,~~~|2\re=c_{j,1}^\dagger|0\re\,,~~~
  |3\re=c_{j,2}^\dagger|0\re\,,~~~ |4\re=c_{j,3}^\dagger|0\re,\no\\
&&|5\re=c_{j,1}^\dagger c_{j,2}^\dagger|0\re\,,~~~ 
  |6\re=c_{j,1}^\dagger c_{j,3}^\dagger|0\re\,,~~~ 
  |7\re=c_{j,2}^\dagger c_{j,3}^\dagger|0\re\,,~~~ 
  |8\re=c_{j,1}^\dagger c_{j,2}^\dagger c_{j,3}^\dagger|0\re\,.\label{choice}
\eea
Then by (\ref{projector1}, \ref{projector2}), (\ref{basis1},
\ref{basis2}), (\ref{dual1}, \ref{dual2}) and (\ref{choice}),
and after tedious but straightforward manipulation,
one gets, up to a constant, 
\beq
H_{j,j+1}(g,\k)=-\frac{2\sinh\k g}{\k}
   \;H^{\rm R}_{j,j+1}(g,q=e^{-\k}). 
\eeq
This identity also shows that $H(g,\k)$
commutes with the generators (\ref{matrix}) of $U_q[gl(3|1)]$, 
since the R-matrix $\check{R}(u)$ is a $U_q[gl(3|1)]$ invariant. 

We now solve the system by means of the coordinate space 
Bethe ansatz technique. We assume the following wavefunction
\beq
\psi_{\a_1,\cdots,\a_N}(x_1,\cdots,x_N)=\sum_P\e_P\,A_{\a_{Q_1},\cdots,
  \a_{Q_N}}(k_{P_{Q_1}},\cdots,k_{P_{Q_N}})\,exp\lt(i\sum_{j=1}^N
  k_{P_j}x_j\rt),
\eeq
where $Q$ is the permutation of the $N$ particles such that
$1\leq x_{Q_1}\leq\cdots\leq x_{Q_N}\leq L$. Denote $X_Q=\{x_{Q_1}\leq
\cdots\leq x_{Q_N}\}$. The coefficients $A_{\a_{Q_1},\cdots,\a_{Q_N}}
(k_{P_{Q_1}},\cdots,k_{P_{Q_N}})$ from regions other than $X_Q$ are
connected with each other by elements of two-particle S-matrix:
\bea
S_{ij}(k_i,k_j)^{aa}_{aa}&=&1,~~~ a=1,2,3,\no\\
S_{ij}(k_i,k_j)^{ab}_{ab}&=&\frac {\sin (\l_i-\l_j)}
  {\sin (\l_i-\l_j-i\kappa)},~~~
  a \neq b,~a,b=1,2,3,\no\\
S_{ij}(k_i,k_j)^{ab}_{ba}&=&e^{i {\rm sign} (a-b)(\l_i-\l_j)}
\frac {\sin i\kappa}{\sin (\l_i-\l_j-i\kappa)},~~~ a,b=1,2,3,
\eea
where $\l_j$ are suitable particle
rapidities related to the quasi-momenta $k_j$ of the electrons by
\beq
k(\l)=
2 \arctan (\coth c \tan \l),
\eeq
where the parameter $c$ is defined by
\beq
c= \frac {1}{4} \lt\{ \ln [\frac {\sinh \frac {1}{2}(\eta +\kappa)}
 {\sinh \frac {1}{2}(\eta -\kappa)}]-\kappa \rt\}.
\eeq
The periodicity condition for the system on the finite interval $(0,L)$
results in the Bethe equations for the set of $N$ momenta $k_j$
: $exp(ik_jL)=T_j,~j=1,\cdots,N$, where
\beq
T_j=S_{j,j+1}(k_j,k_{j+1})\cdots S_{j,N}(k_j,k_N)S_{j,1}(k_j,k_1)\cdots
    S_{j,j-1}(k_j,k_{j-1}), ~~~~j=1,\cdots,N.
\eeq
The meaning of $T_j$ is the scattering matrix of the $j$-th particle on
the other $(N-1)$ particles. So now the problem is to diagonalize $T_j$
to arrive at a system of scalar equations. It can be shown that
$T_j=\tau(\l=k_j)$, where 
\beq
\tau(\l)=tr_0\lt[S_{0,1}(\l-k_1)\cdots S_{0,N}(\l-k_N)\rt]
\eeq
is the transfer matrix of the inhomogeneous $U_q[gl(3)]$-spin
magnet of $N$ sites.
The commutativity of the transfer matrix for different values of the
spectral parameter $\l$ implies that $T_j,~j=1,\cdots,N$ can be
diagonalized simultaneously.  The
Bethe ansatz equations are written in terms of the rapidities
$\L_\s^{(1)},~\L_\s^{(2)}$ and $\l_\s$
\bea
e^{ik_j L}&=&\prod_{\s=1}^{M_1}\frac{\sin (\l_j-\L^{(1)}_\s-i\kappa/2)}
      {\sin (\l_j-\L^{(1)}_\s+i\kappa/2)},\no\\
\prod_{j=1}^N\frac{\sin (\L^{(1)}_\s-\l_j+i\kappa/2)}{\sin 
(\L^{(1)}_\s-\l_j-i\kappa/2)}&=&
   -\prod_{\rho=1}^{M_1}\frac{\sin (\L^{(1)}_\s-\L^{(1)}_\rho+i\kappa)}
   {\sin (\L^{(1)}_\s-\L^{(1)}_\rho-i\kappa)}
   \prod_{\rho=1}^{M_2}\frac{\sin (\L^{(1)}_\s-\L^{(2)}_\rho-i\kappa/2)}
   {\sin (\L^{(1)}_\s-\L^{(2)}_\rho+i\kappa/2)},~~~\s=1,\cdots,M_1,\no\\
\prod_{\rho=1}^{M_1}\frac{\sin (\L^{(2)}_\g-\L^{(1)}_\rho+i\kappa/2)}
   {\sin (\L^{(2)}_\g-\L^{(1)}_\rho-i\kappa/2)}&=&-
   \prod_{\rho=1}^{M_2}\frac{\sin (\L^{(2)}_\g-\L^{(2)}_\rho+i\kappa)}
   {\sin (\L^{(2)}_\g-\L^{(2)}_\rho-i\kappa)},~~~\g=1,\cdots,M_2,\label{Bethe-ansatz}
\eea
The energy of the system in the state corresponding to the sets of
solutions $\{\L_\s^{(1)},~\L_\s^{(2)}\}$ and 
$\{\l_\s\}$ is (up to an additive constant, which we drop)
$E=-2\sum_{j=1}^N\cos k_j$.

To summarize, we have presented a new two-parameter integrable model
which is an eight-state supersymmetric electron model with correlated
single-particle and pair hoppings as well as uncorrelated
triple-particle hopping.
We have solved the model by the coordinate
Bethe ansatz method and derived the Bethe ansatz equations. 
There are many problems remained to be done for this new model. One of
them is to incorporate integrable boundary conditions into the model.
We hope to report
results on this aspect in future publications.

\vskip.3in
Y.-Z.Z. is supported by the QEII Fellowship Grant from 
Australian Research Council.



\begin{thebibliography}{99}
\bibitem {Ess94} F.H.L. Essler, V.E. Korepin, {\it Exactly solvable
   models of strongly correlated electrons}, World Scientific, 1994
\bibitem{Kaw90} N. Kawakami, S.-K. Yang, Phys. Lett. {\bf A148}
   (1990) 359;\\
   H. Frahm, V.E. Korepin, Phys. Rev. {\bf B42} (1990) 10553.
\bibitem{Bar95} R.Z. Bariev, A. Kl\"umper, J. Zittartz, Europhys. Lett.
   {\bf 32} (1995) 85;\\
   M.D. Gould, K.E. Hibberd, J.R. Links, Y.-Z. Zhang,
   Phys. Lett. {\bf A212} (1996) 156.
\bibitem{Gou97} M.D. Gould, Y.-Z. Zhang, H.-Q. Zhou, preprint
   cond-mat/9709129.
\bibitem{Bra95} A.J. Bracken, M.D. Gould, J.R. Links, Y.-Z. Zhang,
   Phys. Rev. Lett. {\bf 74} (1995) 2768.
\bibitem{Bed95} G. Bed\"urftig, H. Frahm, J. Phys. {\bf A:} Math.
   Gen. {\bf 28} (1995) 4453;\\
   P.B. Ramos, M.J. Martins, Nucl. Phys. {\bf B474}
   (1996) 678;\\
   M.P. Pfannm\"uller, H. Frahm, Nucl. Phys. {\bf B479}
   (1996) 575;\\
   K.E. Hibberd, M.D. Gould, J.R. Links, Phys. Rev. {\bf B54} (1996)
   8430.
\bibitem{Zha97} Y.-Z. Zhang, H.-Q. Zhou, preprint cond-mat/9711238.
\bibitem{Arn97} D. Arnaudon, preprint physics/9711001.
\bibitem{Bar91} R.Z. Bariev, J. Phys. {\bf A:} Math. Gen. {\bf 24}
   (1991) L919.
\bibitem{Zho97} H.-Q. Zhou, D.-M. Tong,
   Phys. Lett. {\bf A232} (1997) 377. 

\end{thebibliography}
\end{document}